\documentclass[12pt,twoside,pointlessnumbers,smallheadings]{article}
\usepackage{amsfonts}
\usepackage{indentfirst}
\input psfig.sty

\newcommand{\psp}{\psi^{\prime}}
\newcommand{\psip}{\psi(2S)}
\newcommand{\pspp}{\psi^{\prime \prime}}

\newcommand{\jpsi}{J/\psi}
\newcommand{\EE}{e^+e^-}
\newcommand{\MM}{\mu^+\mu^-}

\newcommand{\GG}{\gamma\gamma}
\newcommand{\beq}{\begin{equation}}
\newcommand{\eeq}{\end{equation}}
\def\eref#1{(\ref{#1})}
\def\Journal#1#2#3#4{{#1} {\bf #2}, #3 (#4)}

\def\IJMPA{Int. J. Mod. Phys. A}

\def\NIM{Nucl. Instrum. Methods}

\def\NPB{Nucl. Phys. B}

\def\PLB{Phys. Lett. B}

\def\PRL{Phys. Rev. Lett.}
\def\PRD{Phys. Rev. D}

\def\ZPC{Z. Phys. C}

\def\HEPNP{HEP \& NP}

\begin{document}

\title{On resonance parameter measurement and luminosity determination
at $\EE$ collider\thanks{Supported by National Natural Science
Foundation of China (10491303, 10775412,10825524), Major State Basic
Research Development Program (2009CB825200, 2009CB825206), Knowledge
Innovation Project of The Chinese Academy of Sciences
(KJCX2-YW-N29), Research and Development Project of Important
Scientific Equipment of CAS (H7292330S7).} }

\author{P.~Wang$^{1}$\footnote{E-mail:wangp@mail.ihep.ac.cn},
Y.S.~Zhu$^{1}$\footnote{E-mail:zhuys@mail.ihep.ac.cn},
X.H.~Mo$^{1}$\footnote{E-mail:moxh@mail.ihep.ac.cn},\\
{\small 1 (Institute of High Energy Physics, CAS, Beijing 100049,
China )} }
% \\
%{\small 2 (Tsinghua University, Beijing 100084, China )}   \\
%{\small 3 (Budker Institute of Nuclear Physics, Novosibirsk 630090,Russia )} }

\date{\today}
\maketitle

\begin{center}
\begin{minipage}{15cm}
{\small {\bf Abstract} \hskip 0.25cm Expounded are the parameter
measurement for narrow resonance and determination of corresponding
luminosity at $\EE$ collider. The detailed theoretical formulas are
compiled and the crucial experimental effects on observed cross
section are taken into account. For luminosity determination, the
iteration method is put forth which is mainly used to separate the
interference effect between resonance and non-resonance decays.

{\bf Key words} \hskip 0.25cm  resonance parameter, luminosity,
$\EE$ collider }

%{\bf PACS} \hskip 0.25cm  07.89.+b,  28.41.Qb,  29.40.Wk}
\end{minipage}
\end{center}

%%%%%%%%%%%%%%%%%%%%%%%%%%%%%%%%%%%%%%%%%%%%%%%%%%%%%%%%%%%%%%%%%%%%%%%
\section{Introduction}
%%%%%%%%%%%%%%%%%%%%%%%%%%%%%%%%%%%%%%%%%%%%%%%%%%%%%%%%%%%%%%%%%%%%%%%
Resonance are a special kind of particles and their study is of
great interest and importance in the domain of elementary particle
physics. Measurements of resonance parameters, such as the mass
($M_R$), total decay width ($\Gamma_t$), partial decay width of
final state $f$ ($\Gamma_f$, where $f=e,\mu,\tau$ indicating the
$e$-pair, $\mu$-pair, and $\tau$-pair final states, respectively),
and corresponding branching ratios are fundamental work for high
energy experimental physics. For resonances of $1^{--}$ charmonium
and bottomnium, such as $\jpsi$, $\psp$, $\pspp$, $\cdots$,
$\Upsilon(nS), (n=1,2,3,4, \cdots)$ $etc$, their resonance parameters can
be measured by scan experiment at $\EE$
colliders~\cite{Augustin}-\cite{pdg08}, which is one of basic
approaches to understand resonances in collision experiment.

For the experiment using scan method, data are taken at
several different energy points in the vicinity of the resonance to
be measured. The minimization technique is usually applied on the
estimator which constructed by the difference between the measured
number of events and the expected number of events. The latter can
be obtained by theoretical calculation and Monte carlo simulation.
Specially, the expected number of events for certain final state $f$
at the point with center-of-mass energy $W$ can be obtained by the
expression \beq N_f (W,\vec{\eta}) = L(W) \cdot \sigma^{obs}_f
(W,\vec{\eta})~, \label{expnmb} \eeq where $\vec{\eta}$ is the
parameter vector which contains the information of resonance
parameters. $\sigma^{obs}$ is the experimentally observed cross
section (the detailed description refer to section~\ref{skn_expskn})
which is the synthetic cross section including resonance part,
continuum part, and their interference; and also incorporating the
effect due to experiment efficiency. $L$ is the luminosity which can
be acquired through several approaches.
%%Anyway, as far as the physics analysis is concerned, the luminosity is
%%often determined by some physics processes.

\begin{figure}[bthp]
\centerline{\psfig{file=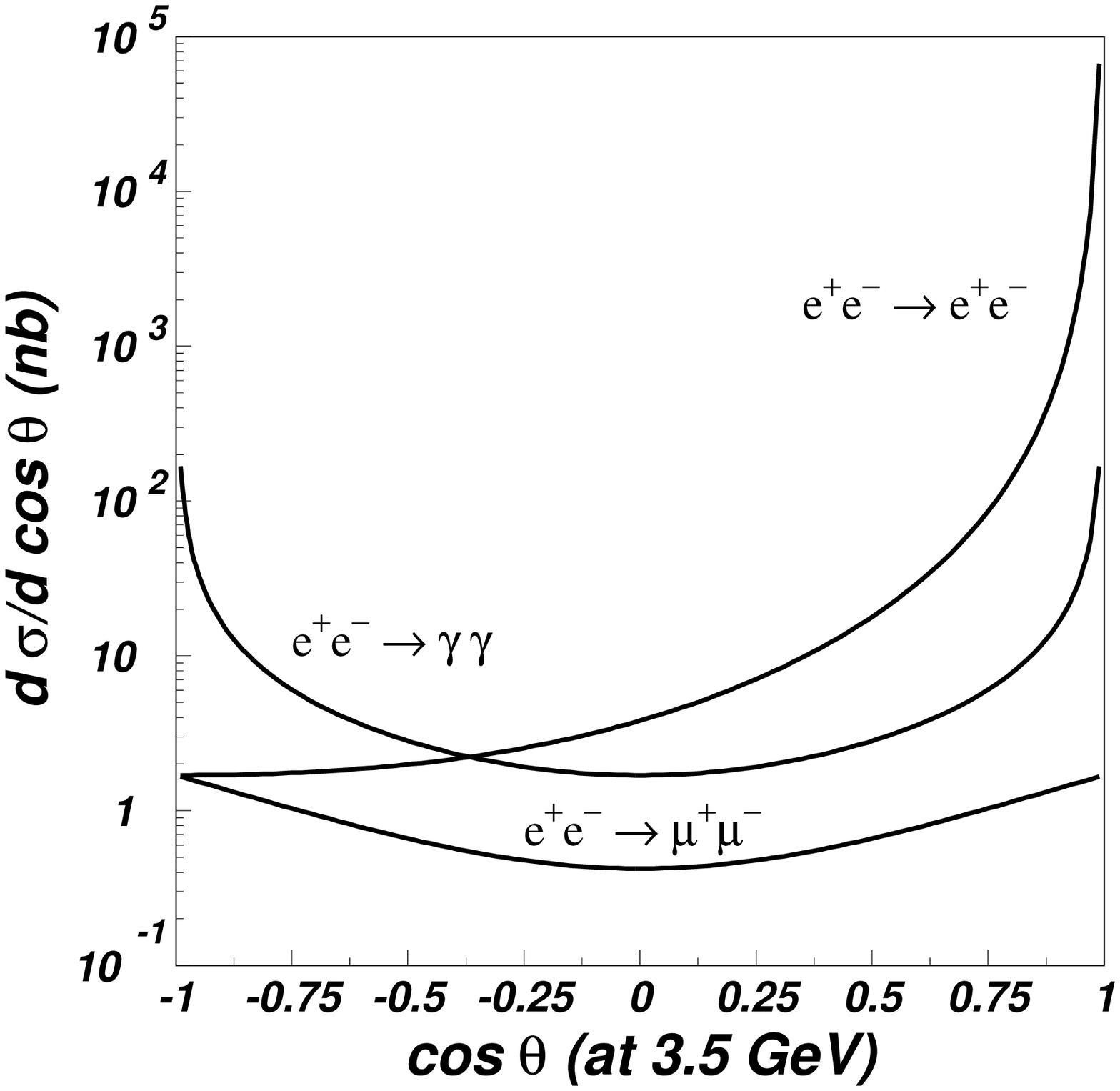,height=8.cm,width=7.5cm}}
\caption{\label{qedxct} Differential cross sections for three QED
processes: $\EE \to \EE$, $\EE \to \MM$, and $\EE \to \GG$. The
center-of-mass energy is 3.5 GeV.}
\end{figure}

In principle any detectable process can be used for luminosity
measurement. However in order to achieve high precision, one often
selects the process which has larger cross section and salient
characteristic topology experimentally with accurate theoretical
calculation of the cross section. From these point of view, the QED
processes such as $\EE \to \EE$, $\EE \to \MM$, and $\EE \to \GG$,
are most often adopted for luminosity measurements~\cite{huanggs}.
The lowest order of differential cross sections for these processes
are shown in Fig.~\ref{qedxct}. Experimentally, the response of the
detector to each of these reactions is quite distinct: efficiencies
rely on the charged particle tracking ($\EE$), calorimetry ($\EE$
and $\GG$), muon counter ($\MM$), and trigger algorithms. The
expected theoretical cross sections are calculable in quantum
electrodynamics; weak interaction effects are negligible for charm
and B-factories at the level of 1 per mill.~\cite{ylbkbes3}.

For luminosity measurement, the interference effect in the vicinity
of resonance peaks has to be treated with great care\footnote{For
$\GG$ final state, since only the continuum process exists, there is
no interference dilemma and luminosity measurement is simple.}. Such
effect not only distorts the cross section in the peak region but
also shifts the resonance peak position. Especially, when the cross
sections of resonance and non-resonance processes are compatible,
the interference effect are too prominent to be neglected. In such
circumstance, when we consider how to determine the luminosity, we come
across a dilemma. On one hand, to determine the luminosity we must
subtract the contributions due to resonances and corresponding
interference effect. This can be realized by correct determination
of resonance parameters. On the other hand, the measurement of
resonance parameters depends on the accurate determination of
luminosity. That is to say the measurement of resonance parameter
and determination of luminosity are the cause-consequence
interdependence. To resolve such an intertwist issue, we recourse to
an iteration approach which will be expounded in
section~\ref{skn_fitdt}. Before that, in section~\ref{skn_resxkn}
and~\ref{skn_expskn}, presented are the formulas for experimentally
observed cross section which take into account various experimental
effects at $\EE$ collider such as vacuum polarization, initial
radiative correction, and beam energy spread.

\section{Cross Section for $1^{--}$ resonance}\label{skn_resxkn}
In this section, we discuss in detail the experimental corrections
on cross section and provide the analytic expressions for
calculation of experimentally observed cross section.

\subsection{Experimental corrections}

The cross section of the resonance process \beq e^{+} e^{-}
\rightarrow \mbox{Res.} \rightarrow f~, \label{resprocess} \eeq
where $f$ denotes a certain kind of final state, is described by the
Breit-Wigner formula
\begin{equation}
\sigma_{BW}(W)=\frac{12 \pi \cdot \Gamma_{e} \Gamma_{f} }
                   {(W^2-M^2)^2+\Gamma^2 M^2} ,
\label{bwxct}
\end{equation}
where $W$ is the center-of-mass energy, $\Gamma_{e}$ and
$\Gamma_{f}$ are the widths of the resonance decaying into $e^{+} e^{-}$
and $f$, $\Gamma$ and $M$ are the total width and mass of resonance.
Taking the initial state radiative (ISR) correction into
consideration, the cross section becomes \cite{Kuraev85}
\begin{equation}
\sigma_{r.c.} (W)=\int \limits_{0}^{x_m} dx F(x,s) \frac{1}{|1-\Pi
(s(1-x))|^2} \sigma_{BW}(s(1-x)) , \label{isrxct}
\end{equation}
where $s=W^2$, $x_m=1-s^{\prime}/s$, $\sqrt{s^{\prime}}$ is the
experimentally required minimum invariant mass of the final state
$f$ after losing energy due to multi-photon emission; $F(x,s)$ has
been calculated in many
references~\cite{Kuraev85,Altarelli,Nicrosini,Berends} and $\Pi
(s(1-x))$ is the vacuum polarization factor. The radiative
correction in the final states are usually not
considered~\cite{tsai,berends78}. The reasons are twofold. In the
first place, the hadronic final system is very complicated and since
the radiative corrections depend upon the details of how the
experiment is done, it is difficult to give a general,
model-independent prescription for them. The second reason is that
our understanding of the hadronic problem is so crude that there is
no need to worry about the electromagnetic corrections\footnote{In
any case, if we find later on that it is necessary to do radiative
corrections to the hadronic states for some specific problem, we can
do the calculation then, because the initial state radiative
corrections and final state radiative corrections can be decoupled
to a large extent.}.

The $e^+e^-$ colliders have finite beam energy spread. The beam
energy spread function $G(W,W')$ is usually a Gaussian distribution:
\begin{equation}
G(W,W^{\prime})=\frac{1}{\sqrt{2 \pi} \Delta}
             e^{ -\frac{(W-W^{\prime})^2}{2 {\Delta}^2} },
\end{equation}
where $\Delta$ is the standard deviation of the Gaussian
distribution. It varies with the beam energy of the collider. For
narrow resonances such as $\jpsi$ and $\psp$, $\Delta$ is usually
much wider than the resonance intrinsic width. Therefore the
beam-spreaded resonance cross section is the radiatively corrected
Breit-Wigner cross section folded with the energy spread function:
\begin{equation}
\sigma_{G} (W)=\int \limits_{0}^{\infty}
        dW^{\prime} \sigma_{r.c.} (W^{\prime}) G(W^{\prime},W),
\label{obsxct}
\end{equation}
where $\sigma_{r.c.}$ is defined by Eq.~\eref{isrxct}.

\begin{figure}[bthp]
\centerline{\psfig{file=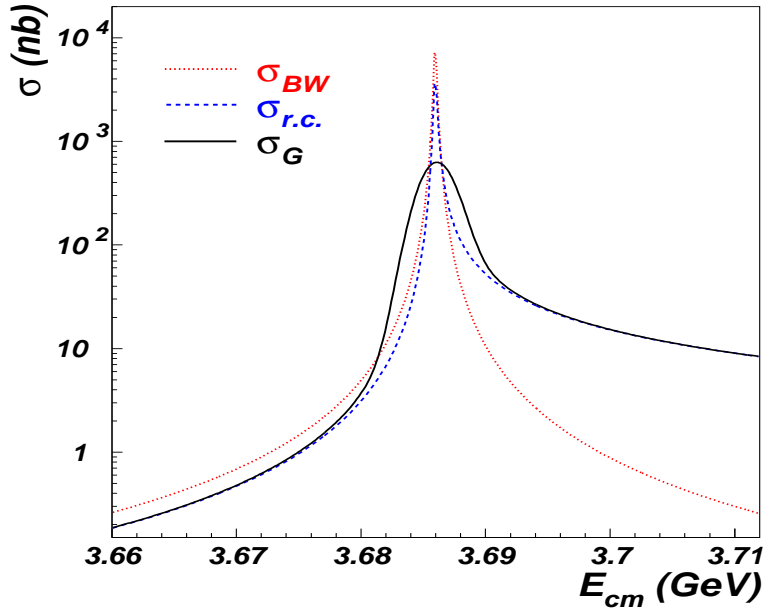,height=8.cm,width=10. cm}}
\caption{\label{cmpbrg}Inclusive hadronic cross sections for
resonance of $\psp$: $\sigma_{BW}$ for Breit-Wigner cross section;
$\sigma_{r.c.}$ the cross section with radiative correction;
$\sigma_{G}$ the beam-spreaded cross section with the energy spread
effect. In the calculation of these cross sections, the following
parameters are used~\cite{pdg08}: $M=3.68596$ GeV, $\Gamma_{t}=300$
keV, $\Gamma_{e}=2.19$ keV, and $\Delta=1.3$ MeV.}
\end{figure}

Take resonance $\psp$ as an example, Fig.~\ref{cmpbrg} displays
three cross sections: the Breit-Wigner cross section of
Eq.~\eref{bwxct}; the cross section after radiative correction by
Eq.~\eref{isrxct}, and the beam-spreaded cross section by
Eq.~\eref{obsxct}. From the three curves in Fig.~\ref{cmpbrg}, it
can be seen that the radiative correction reduces the height of the
resonance. It also shifts the peak position to above the $\psip$
nominal mass. The reduction factor $\rho$ and the shift of the peak
$\Delta\sqrt{s}_{max}$ are approximately expressed
by~\cite{berends89}

\begin{eqnarray}
\rho &=&{\displaystyle \left(\frac{\Gamma}{M}\right)^t
 \cdot (1+ \delta)}, \\
\Delta \sqrt{s}_{max} &=&{\displaystyle \frac{t \pi}{8} \Gamma },
\end{eqnarray}
where $t$ is defined as \beq t = \frac{2 \alpha}{\pi} \left( \ln
\frac{s}{m^2_e} -1 \right), \label{defoft} \eeq with $\alpha$ and
$m_e$ the QED fine structure constant and the mass of electron;
\begin{equation} \delta=\frac{\alpha}{\pi} (\frac{\pi^2}{3} -
\frac{1}{2})+ \frac{3}{4} t + t^2(\frac{9}{32}-\frac{\pi^2}{12})
\label{defdelta}
\end{equation}
Herein the reduction factor $\rho$ is the ratio of the maximum
resonance cross section with radiative correction to that of Born
order.

At the $\psp$ mass, $t \approx 0.0779$ and $\delta \approx 0.06$,
then the reduction factor $\rho \approx 0.51$ and the shift of the
peak $\Delta \sqrt{s}_{max} \approx 9$ keV. The energy spread
further lowers down and shifts the experimentally observed $\psp$
peak. In the case of a collider with $\Delta=1.3$ MeV, the maximum
height of the $\psip$ peak becomes 640 nb, and the position of the
peak is shifted by 0.14 MeV above the $\psp$ nominal mass.

From the example discussed above, the effects due to experimental
corrections on resonance cross section are fairly prominent. For the
continuum, effects are comparatively moderate~\cite{wangp06}. Next,
we will discuss in detail the ISR effect on Breit-Wigner cross
section.

\subsection{Initial state radiative correction}
The ISR correction scheme used by earlier experiments, is based on
the work of Bonneau and Martin~\cite{bonneau} and that of Jackson
and Scharre~\cite{jackson}. The former only calculated to $\alpha^3$
order which is insufficient for resonances; while the latter made
some mistakes~\cite{Alexander88,Alexander89}. The drawbacks due to
the treatment of the radiative correction with these two schemes
were studied for Z in Ref.~\cite{Alexander88} and for narrow
resonances of $\psi$ and $\Upsilon$ families in
Ref.~\cite{Alexander89}. In the eighties of the last century Kuraev and
Fadin treats ISR correction based on the structure function approach
which achieves 0.1\% accuracy~\cite{Kuraev85}. Afterwards such an
approach is extensively used for ISR correction which is also followed
in this paper.

The calculation of $F(x,s)$ is summarized in Ref~\cite{berends89}.
But for the discussions on resonances in this paper, a different
form derived in Ref~\cite{chenfz} is more useful:

\beq F(x,s)= x^{t-1}t \cdot ( 1+\delta )
+x^t(-t-\frac{t^2}{4})+x^{t+1}(\frac{t}{2}-\frac{3}{8}t^2)+{\cal O}
 (x^{t+2}t^2)~.
\label{deffcnxs}\eeq Here the conversion of soft photons
into real $e^+e^-$ pairs is included.
Notice that $x<1$ and $t<1$, so the omitted terms
in the above equation are small quantities. Then using
equality~\cite{Cahn}
\begin{equation}
\int \limits^{\infty}_0 \frac{\nu x^{\nu -1}dx}{x^2+2ax \cos\beta
+a^2} =a^{\nu -2} \cdot \frac{\pi\nu \sin[\beta(1-\nu)]}{\sin \beta
\sin \pi \nu}~ (\mbox{For  } \nu <2 )~,
\end{equation}
It can be obtained finally~\cite{chenfz}
$$\sigma_{r.c.}(s)= \frac{12\pi\Gamma^{exp}_e\Gamma_f}{s^2}
\cdot \left\{ (1+\delta)t \cdot \frac{}{}\right.$$
$$\left[ \frac{1}{t} \cdot a^{t -2} \phi(\cos\beta,t)
+\frac{1}{t-2} +\frac{2(s-M^2)}{(t-3)s} +\frac{3(s-M^2)^2-M^2
\Gamma^2}{(t-4)s^2} \right]$$
$$ -t\left(1+\frac{t}{4}\right) \cdot $$
$$\left[ \frac{1}{(t+1)} \cdot a^{t -1} \phi(\cos\beta,t+1)
+ \frac{1}{t-1} +\frac{2(s-M^2)}{(t-2)s} +\frac{3(s-M^2)^2-M^2
\Gamma^2}{(t-3)s^2} \right]$$
$$ +t \left(\frac{1}{2}-\frac{3}{8}t \right) \cdot $$
\begin{equation}
\mbox{} \hskip -0.5cm \left.\left[ \frac{1}{2} \ln \frac{1+2 a \cos
\beta +a^2} {a^2} - \mbox{ctg} \beta \cdot \left( \arctan \frac{1+ a
\cos\beta }{a \sin \beta} -\frac{\pi}{2} + \beta \right)
\right]\right\}~, \label{xctapprox}
\end{equation}
%where $\delta$ defined in Eq.~\eref{defdelta}.
where
\begin{eqnarray}
   a^2            &=&\left(1-\frac{M^2}{s}\right)^2 + \frac{M^2 \Gamma^2 }{s^2}
                     \hskip 0.5cm (a>0) \label{asqdef}\\
\cos \beta       &=&\frac{1}{a} \cdot \left(\frac{M^2}{s}-1\right) \label{cosdef}\\
\phi(\cos\beta,y)&=&\frac{\pi y \sin[\beta(1-y)]}
                          {\sin \beta \sin \pi y}  \label{phidef}
\end{eqnarray}

In fact, a more simplified formula can be used , viz.
%$${\sigma}_{r.c.} (s)
%=\frac{12 \pi \Gamma^{exp}_e \Gamma_f }{W^4} (1+\delta_{vac}(s))
%\cdot$$
\begin{equation}
{\sigma}_{r.c.} (s)=\frac{12 \pi \Gamma^{exp}_e \Gamma_f }{W^4}
\left\{ (1+\delta) a^{t-2} \phi(\cos\beta,t)+ (-t+\frac{3}{4} t^2)
a^{t-1} \phi(\cos\beta,t+1) \right\}~. \label{xctsimple}
\end{equation}

For resonances of $\psi$ and $\Upsilon$ families, the accuracy of
expression~\eref{xctapprox} is better than 0.1\% while that of
expression~\eref{xctsimple} is better than 0.2\% over a sufficient
large energy range around the resonances which is usually scanned by the
experiments. Therefore even the
latter is accurate enough to be used for present data
fit~\cite{chenfz}.

One remark is in order here. As aforementioned, in Eq.~\eref{bwxct},
$\Gamma_{e}$ and $\Gamma_{f}$ are the partial widths of the $\EE$
mode and the final state $f$ (here $f$ usually indicates the
hadronic final state) respectively. Here $\Gamma_{e}$ describes the
coupling strength of the resonance to $\EE$ through a virtual
photon. For example, in potential model, $\Gamma_{e}$ is related to
the wave function at the origin $\psi(0)$ in the way
$$%%\begin{equation}
\Gamma_{ee} = \frac{4\alpha^2 Q_q^2 |\psi(0)|^2}{M^2}~,
$$%%\end{equation}
where $Q_q$ is the charge carried by the quark in the quarkonium and
$\alpha$ is the QED fine structure constant. Since the decay of a
quarkonium $1^{--}$ state to $\EE$ pair is through a virtual photon,
there is always vacuum polarization associated with this process. So
the experimentally observed $\EE$ partial width, denoted explicitly
as $\Gamma_{e}^{exp}$, is related to $\Gamma_{e}$ by the expression
$$%%\begin{equation}
\Gamma _{e}^{exp}=\frac{\Gamma_{ee}}{|1-\Pi(M^2)|^2}~.
%%\label{gee}
$$%%\end{equation}
This is the convention of Ref.~\cite{tsai,Alexander89} which is
adopted by PDG. In this convention $\Gamma_{e}$ means
$\Gamma_{e}^{exp}$. So in the above discussion, the factor of vacuum
polarization has been absorbed into the partial decay width of $\EE$
final state, as in Eqs.~\eref{xctapprox} and \eref{xctsimple}.
However, when the leptonic decay is concerned, only one polarization
factor can be absorbed into $\Gamma_{e}$ or $\Gamma_{\mu}$.
Therefore in the following formulas, the vacuum polarization factor
will be given explicitly.

\subsection{Vacuum polarization}
A pedagogical description on the calculation of vacuum polarization can
be found in many textbooks on quantum electrodynamic, e.g.
Ref.~\cite{Greiner}. In this section, we merely
collected the formulas for the following usage.

In the actually calculation, the polarization factor is often expressed
as $1+ \delta_{vac}$ with relation \beq 1+ \delta_{vac} =
\frac{1}{|1-\Pi(s)|^2}~.\eeq\footnote{The definition of $\Pi(s)$
in the literature varries by a minus sign, e.g. between
Ref.~\cite{Kuraev85} and \cite{Berends76}.}
According to the calculation of field
theory, \begin{equation}
\begin{array}{lcl}
 \delta_{vac}&=& -2 \cdot Re \Pi (s)
  \hskip 4.2cm \makebox[0.8cm]{($s$ \mbox{channel} )}\\
             & & -2 \cdot Re \Pi (t)
  \hskip 4.3cm \makebox[0.8cm]{($t$ \mbox{channel} )}\\
             & & - Re \Pi (s)\cdot Re \Pi (t)
  \hskip 2.5cm \makebox[2.4cm]{~~~(~~~\mbox{interference bwteen} $s$ \mbox{ and } $t$)}
\end{array}
\end{equation}

\subsubsection{Leptonic part}
When $s>4 m_l^2$, $m_l$ is the mass of lepton ($l=e,\mu,\tau$)
\begin{equation}
 Re \Pi(s,m_l^2)=\frac{\alpha}{\pi}\left[\frac{8}{9}-\frac{a_l^2}{3}
+a_l \cdot \left(\frac{1}{2}-\frac{a_l^2}{6} \right) \cdot \ln b_l
\right]~,
\end{equation}
where
\begin{equation} a_l=\left(1-\frac{4 m_l^2}{s}\right)^{\frac{1}{2}}, \hskip
1 cm b_l=\frac{1-a_l}{1+a_l}~.
\end{equation}
When $s<4 m_l^2$
\begin{equation}
 Re \Pi (s,m_l^2)=\frac{\alpha}{\pi}\left[ \frac{8}{9}+\frac{a_l^2}{3}
-2a_l \cdot \left(\frac{1}{2}+\frac{a_l^2}{6}\right) \cdot
\cot^{-1}(a_l) \right]
\end{equation}
where
\begin{equation}
a_l=\left(\frac{4 m_l^2}{s}-1\right)^{\frac{1}{2}}~.
\end{equation}
For $t<0$, $t=-s \cdot \sin^2\frac{\theta}{2}$($\theta$: polar
angle), so $Re \Pi(t,m_l^2)$ is the function of the center-of-mass
energy $E_{cm}$ and scattering angle $\theta$.
\begin{equation}
Re \Pi(t,m_l^2)=\frac{\alpha}{\pi}\left[\frac{8}{9}-\frac{a_l^2}{3}
+a_l \cdot \left(\frac{1}{2}-\frac{a_l^2}{6} \right) \cdot \ln(-
b_l) \right]
\end{equation}
where
\begin{equation}
a_l=\left(1-\frac{4 m_l^2}{t}\right)^{\frac{1}{2}}, \hskip 1 cm
b_l=\frac{1-a_l}{1+a_l}~.
\end{equation}
%%Reference : F.A.Berends and R.Gastmans in ``Radiative corrections in
%%$e^+ e^-$ collision'' Eq(3.19) $\&$ Eq(3.20) and Y.s.Tsa :SLAC-PUB-3129(1983) Eq(2.7). \\
%%Notice the definitions of $\Pi(s)$ in the above two papers have
%%opposite signs.Here Berends' convention is used.\\

\subsubsection{Hadronic part}
For $s$ channel~\cite{Berends76} \mbox{} \hskip 2cm \beq
\begin{array}{l} Re\Pi_{h}(s)= {\displaystyle
\frac{-3s}{\alpha}\sum\limits_{resonances} \frac{\Gamma_{e}}{M}
\cdot \frac{s-M^2}{M^2 \Gamma^2_{tot}+(s-M^2)^2} \hskip 0.5cm
\makebox[2.0cm]{~~(\mbox{ resonance part})} } \\
  {\displaystyle -\frac{\alpha}{3 \pi} R(s_1) \ln \left| \frac{s-s_1}{s_1}\right|
 + \frac{s}{4 \pi^2 \alpha} \int^{s_1}_{4 m^2_{\pi}}
\frac{\sigma_h (s^{\prime})-\sigma_h (s)}{s^{\prime}-s}
ds^{\prime}}\\
  {\displaystyle + \frac{s \cdot \sigma_h(s)}{4 \pi^2 \alpha} \ln \left|
\frac{s_1-s}{4 m^2_{\pi} - s} \right|~. \hskip 1.5cm
\makebox[4.0cm]{~~~~(\mbox{  continuum part, viz. $R$-value part})} }\\
\end{array}
\label{vacfac.hs} \eeq In the above equation, the summation includes
all $1^{--}$ resonances, such as $\rho$, $\omega$, $\phi$, $\jpsi$,
$\psp$, $\Upsilon$ and so forth. Where $R(s_1)$ is $R$-value at
$s_1$\footnote{ $s_1$ is a large energy scale, e.g. in the
program of Berends $s_1=(9.5^2+10)$ $GeV^2$.}. In the second term of
Eq.~\eref{vacfac.hs}, $R(s_1)$ can be expressed through
$\sigma_h(s_1)$ as
\begin{equation} - \frac{s_1 \cdot
\sigma_h(s_1)}{4 \pi^2 \alpha} \ln \left| \frac{s-s_1}{s_1}
\right|~,
\end{equation}
where $\sigma_h(s_1)$ indicates the hadronic cross section produced
in $\EE$ collider at $s=s_1$. When $s< 4 m^2_{\pi}$,
$\sigma_h(s)=0$.

For $t$ channel, the variable of $s$ should be changed into $t$ in
Eq.~\eref{vacfac.hs}, but the value $s_1$ should be kept the same.

It should be noticed that when a resonance is to be fit, the
contribution from itself is not included in the summation of
Eq.~\eref{vacfac.hs} to calculate of the vacuum polarization. In
this case, the vacuum polarization is always a smooth function in
the vicinity of the resonance and can be treated as a constance.
Always the calculation program is readily
available~\cite{Jegerlehner}.

\section{Experimentally observed cross sections}\label{skn_expskn}
In $e^+e^-$ colliding beam experiment, for a final state $f$, besides
the decays from resonance which are produced by $\EE$ annihilation
(refer to Eq.~\eref{resprocess}), most often the process
\begin{equation}
e^+e^- \rightarrow \gamma^*  \rightarrow f
\end{equation}
produces the same final state simultaneously, which is
indistinguishable from that due to resonance decays. So for the
final state produced in $e^+e^-$ experiment, it generally composes
of three parts: the resonance, the continuum, and the interference
between them.

In addition, another factor which should be considered is the
experimental acceptance which actually includes trigger efficiency,
reconstruction efficiency, and selection efficiency. For the last
term, it includes the geometry efficiency implicitly (through the
sub-detector coverage) or explicitly (by applying certain angle
cut).

Therefore, the so-call experimentally observed cross sections denote
total cross sections which include interference effect and
acceptance. In the content that follows, we will present the
detailed formulas of cross section for inclusive hadronic final
state\footnote{For exclusive process, form factor has usually to be
taken into account, see the details in
Refs.~\cite{Wang03,Yuan03,Wang03b}}, $\MM$ final state and $\EE$
final state respectively.

\subsection{Hadronic final state}
The experimentally observed cross section for inclusive hadronic
final state is as follows \beq \sigma_h^{obs}(W)=A_h^R
\sigma_h^R(W)+ A_h^C \sigma_h^C(W)~. \label{xctobshad} \eeq
\footnote{For the inclusive hadronic final state, $\sigma^I \propto
\sum Q_i$, where $Q_i$ is the charge of the quark flavor. So for
charmonium region, $3\sigma^I \approx 0$ after summation of $u$, $d$
and $s$ quarks.} Herein the observed cross section has taken into
account the effect due to the energy spread, that is \beq
\sigma_{f}^{R,C,I} (W)=\int \limits_{0}^{\infty} dW^{\prime}
G(W,W^{\prime}) \tilde{\sigma}_{f}^{R,C,I} (W^{\prime})~,
\label{xctexpobs} \eeq where $A_h^R$ and $A_h^C$ are acceptances for
resonance and continuum hadronic events
respectively~\cite{bespspscan}; $f$ indicates the final state which
can be inclusive hadron ($h$), $\mu$-pair ($\mu$), and $e$-pair
($e$) respectively; $R$, $C$, and $I$ denote resonance, continuum,
and interference respectively. In this section, $\tilde{\sigma}$
denotes the ISR corrected cross section (which is denoted by
$\sigma_{r.c.}$ in the previous section).

In Eq.~\eref{xctobshad}, the resonance cross section has been given
in Eq.~\eref{xctsimple}; for non-resonance part:
\begin{equation}
\sigma_h^C=\tilde{R} \sigma_{\mu}^0 (QED)=\tilde{R} \frac{4 \pi
\alpha^2}{3W^2} =\tilde{R} \frac{86.85 (\mbox{nb})}{W^2
(\mbox{GeV}^2)}
\end{equation}
where $\tilde{R}$ is the $R$-value with radiative correction, viz.
\begin{equation} \tilde{R} =R\cdot (1+\delta_{rad}) \end{equation}
where $\delta_{rad}$ is radiative correction factor at the
non-resonance region~\cite{besrval,huhm}.

%%%%%%%%%%%%%%%%%%%%%%%%%%%%%%%%%%%%%%%%%%%%%%%%%%%%%%%%%%%%%%%%%%%%%%
\subsection{$\MM$ final state}
The experimentally observed cross section for $\MM$ final state is
\begin{equation}
\sigma_{\mu}^{obs}(W)=A_{\mu}^R \cdot \sigma_{\mu}^R(W)+
A_{\mu}^{C}\cdot \sigma_{\mu}^C(W)+ A_{\mu}^{I}\cdot
\sigma_{\mu}^{I}(W)~.
\end{equation}
As indicated in Eq.~\eref{obsxct} %Eq.~\eref{xctexpobs},
the effect due to the energy spread has been taken into account for
the cross section in $\sigma_{\mu}^{obs}$. The $\tilde{\sigma}$, which
is denoted as $\sigma_{r.c.}(W)$ in Eq.~\eref{obsxct}, can be
approximated analytically by the following form

$$ \tilde{\sigma} (W=\sqrt{s})=(1+\delta_{vac}(s)) \cdot $$
$$\left\{ \frac{4 \pi \alpha^2}{3 W^2}\cdot A_{\mu}^{C} \cdot
\left[1+\frac{t}{2}\cdot \left(2 \ln X_f -\ln(1-X_f)+\frac{3}{2}-X_f
\right) +\frac{\alpha}{\pi}(\frac{\pi^2}{3}-\frac{1}{2}) \right]
\right.$$
$$+C_1 \cdot (1+\delta) \cdot \left[ a^{t-2}\phi(\cos\beta,t)
+t \cdot \left( \frac{X_f^{t-2}}{t-2}
   +\frac{X_f^{t-3}}{t-3} R_2
   +\frac{X_f^{t-4}}{t-4} R_3 \right) \right] $$
$$\mbox{} \hskip -0.4cm
+\left[ -t (1+\delta) C_2 +(-t -\frac{t^2}{4}) C_1 \right] \cdot
\left[ \frac{a^{t-1}}{1+t} \phi(\cos\beta,t+1)
+\frac{X_f^{t-1}}{t-1} +\frac{X_f^{t-2}}{t-2} R_2
+\frac{X_f^{t-3}}{t-3} R_3 \right] $$
$$+\left[ \frac{1}{2} \ln \frac{X_f^2+2 a X_f \cos \beta +a^2}
{a^2} - \mbox{ctg} \beta \cdot \left( \arctan \frac{X_f+a \cos\beta
}{a \sin \beta} -\frac{\pi}{2} + \beta \right) \right] \cdot $$
\begin{equation}
\left. \left[(t+\frac{t^2}{4}) \cdot C_2 +( \frac{t}{2}- \frac{3}{8}
t^2) \cdot C_1 \right] \right\}
\end{equation}
with
\begin{eqnarray}
 R_2&=& \frac{2 (s-M^2)}{s}= -2 a \cos \beta \\
 R_3&=& a^2 ( 4 \cos^2 \beta -1) \\
 X_f&=& 1 - \frac{s_m}{s}
\end{eqnarray}
where $s_m=4E^2_{cut}$, $E_{cut}$ is the energy cut for $\mu$, with
$E_{cut} \ge m_{\mu}$. The definition of the other
variables $C_1$ and $C_2$ read
\begin{eqnarray}
\mbox{} \hskip -1.20cm C_1&=&\left[8 \pi \alpha \cdot
\frac{\sqrt{\Gamma_e \Gamma_{\mu}}}{M} \cdot (s-M^2) \cdot
A_{\mu}^{I} +12 \pi \left(\frac{\Gamma_e \Gamma_{\mu}}{M^2}\right)
\cdot s \cdot
A_{\mu}^{R} \right] \Big/ s^2 ~,\\
%\end{eqnarray}
%\begin{equation}
\mbox{} \hskip -1.20cm C_2&=&\left[8 \pi \alpha \cdot
\frac{\sqrt{\Gamma_e \Gamma_{\mu}}}{M} \cdot A_{\mu}^{I} +12 \pi
\left(\frac{\Gamma_e \Gamma_{\mu}}{M^2}\right) \cdot
A_{\mu}^{R}\right] \Big/ s~.
\end{eqnarray}
In the above expression, the terms with $\sqrt{\Gamma_e
\Gamma_{\mu}}$ indicate the interference part while the terms with
$\Gamma_e \Gamma_{\mu}$ are for the resonance part. The variables
$t$, $\phi$ and $a$ are given in Eqs.~\eref{defoft}, \eref{phidef}
and \eref{asqdef}respectively.

\subsection{$\EE$ final state}
The experimentally observed cross section for $\EE$ final state is
\begin{equation}
\sigma_{e}^{obs}(W,\vartheta)=A_{e}^R(\vartheta) \sigma_{e}^R(W)+
A_{e}^{QED}(\vartheta) \sigma_{e}^C(W)+ A_{e}^{I}(\vartheta)
\sigma_{e}^{Int}(W) \label{xctobsee}
\end{equation}
Since QED cross section of $\EE$ final state is divergent at the
small angle, the acceptance of $A(\vartheta)$ is relevant to the
certain Monte Carlo simulation angle\footnote{It should noted that
the event produced angle $|\cos \vartheta_{prd}|$ ($p$: produce)
must be greater than the event selection angle $|\cos
\vartheta_{sel}|$ ($sel$: selection).}, that is the $\EE$ events are
produced within the scope ($\vartheta \rightarrow \pi -\vartheta$).

The special expressions for the cross section in the above equation are
as follows: \\
1) for resonance
$$\tilde{\sigma}^R (W)=\frac{2 \pi H^2}{s}(1+\delta_{vac}(s)) \cdot $$
\begin{equation}
\left\{ T_{R0}(1+\delta) a^{t-2}\phi(\cos\beta,t)
+[T_{R0}\delta_1+T_{R1} t (1+\delta)]
a^{t-1}\frac{\phi(\cos\beta,t+1)}{1+t} \right\}~;
\end{equation}
2) for continuum or QED process
$$\tilde{\sigma}^{QED} (W)=\frac{2 \pi \alpha^2}{s} (1+\delta_{vac}(s)) \cdot $$
\begin{equation}
\left\{C_{\gamma 0}(1+\delta) x_0^t + \frac{1}{1+t}[C_{\gamma
0}\delta_1 +C_{\gamma 1} t (1+\delta)] x_0^{t+1} \right\}~;
\end{equation}
%\begin{equation}
%+\sigma_{0,s}^{QED}(W) \cdot \delta_{vac}(s) +\sigma_{0,t}^{QED}(W)
%\cdot \delta_{vac}(t) +\sigma_{0,int}^{QED}(W) \cdot
%\frac{\delta_{vac}(s)+\delta_{vac}(t)}{2}
%\end{equation}
3) for the interference
%$$ \tilde{\sigma}^{Int} (W)=\frac{2 \pi \alpha H }{s}
%\sqrt{(1+\delta_{vac}(s))(1+\delta_{vac}(t))} \cdot $$
$$ \tilde{\sigma}^{Int} (W)=\frac{2 \pi \alpha H }{s} \cdot
\left( 1+\frac{\delta_{vac}(s)+\delta_{vac}(t)}{2} \right) \cdot $$
$$\left\{C_{i0}(1+\delta) \cdot \frac{s-M^2}{s} \cdot
a^{t-2}\phi(\cos\beta,t)\right.$$
$$+\left[-C_{i0} t (1+\delta) s +C_{i0} \delta (S-M^2)
+C_{i1} t (1+\delta) (s-M^2) \right] \cdot $$
\begin{equation}
\left. \frac{1}{(1+t)s} \cdot a^{t-1} \phi(\cos\beta,t+1) \right\}~.
\label{eesect}
\end{equation}
The special meaning of parameters in Eq.~\eref{eesect} is
$$H= 3 \frac{\Gamma_e^0}{M}~, \hskip 0.5cm
\delta_1=-t -\frac{t^2}{4}~,  \hskip 0.5cm
 x_0=\frac{2 C_m}{1+C_m}~,$$
with the variables $t$ and $\phi$ given in Eqs.~\eref{defoft} and
\eref{phidef} respectively. In addition,
$$ T_{10}=2 \left(\frac{1+C_m}{1-C_m}-\frac{1-C_m}{1+C_m} \right)
{}\hskip 5.7cm T_{t0}=T_{10}+T_{20}+T_{30}$$
$$ T_{20}=2 \ln \left(\frac{1-C_m}{1+C_m} \right)
{}\hskip 7.0cm T_{i0}=T_{20}+4 T_{30}+T_{40}$$
$$ T_{30}=C_m
{}\hskip 9.3cm T_{R0}=T_{30}+T_{40}+T_{50}$$
$$ T_{40}=-C_m
{}\hskip 9.1cm T_{t1}=T_{11}+T_{21}+T_{31}$$
$$ T_{50}=\frac{1}{12}(6C_m+2C_m^3)
{}\hskip 7.0cm T_{i1}=T_{21}+4 T_{31}+T_{41}$$ \mbox{} \vskip -1cm
$$ T_{11}=- 4 \frac{1-C_m}{1+C_m}
{}\hskip 8cm T_{R1}=T_{31}+T_{41}+T_{51}$$
$$ T_{21}=2 \left(1+ \ln\frac{1-C_m}{1+C_m} \right)
{}\hskip 6.5cm C_{\gamma 0}=T_{t0}+T_{i0}+T_{R0}$$
$$ T_{31}=\frac{1}{2}[2C_m+C_m(1+C_m)-(1+C_m)]
{}\hskip 4cm C_{\gamma 1}=T_{t1}+T_{i1}+T_{R1}$$
$$ T_{41}=\frac{1}{2}[-2C_m-2C_m(1+C_m)+(1+C_m)^2]
{}\hskip 4cm C_{i 0}=2T_{R0}+T_{i0}$$
$$ T_{51}=\frac{1}{12}[6C_m+2C_m^3 +\frac{3}{2}(1+C_m)(6C_m+2C_m^3)
-3(1+C_m)^3] {}\hskip 0.6cm C_{i 1}=2T_{R1}+T_{i1}$$ where
$C_m=|\cos\vartheta_{max}|$, $\vartheta_{max}$ is the largest angle
for event selection, that is $(-C_m <\cos \vartheta < C_m) $.

\section{Fit of experimental data}\label{skn_fitdt}
Assume that the scan data are taken at $n_{pt}$ points with
different energies $W_i$ ($i=1,2,\cdots, n_{pt}$), at each point,
the number of experimentally selected events for final state $f$ is
denoted as $n_f^i$ ($f=h(\mbox{hdaron}),
e(\mbox{e-pair}),\mu(\mbox{$\mu$-pair})$); the theoretically
expected number is calculated by the formula \beq N_f^i
(\vec{\eta})= L^i \cdot \sigma_f^i(\vec{\eta})~, \label{expnmb} \eeq
where $L^i$ is the luminosity measured at energy $W_i$ and the
energy dependence has been denoted simply by the subscribe ``$i$'';
$\sigma$ indicates the observed cross sections given in the previous
section with the superscription ``$obs$'' removed in this section;
$\vec{\eta}$ is the parameter vector which contains all parameters
to be fit, such as resonance mass ($M_R$), total decay width
($\Gamma_t$), partial decay width of final state $f$ ($\Gamma_f$),
energy spread ($\Delta$), and so on. For example, for $\psp$ scan
(assuming $e$-$\mu$ universality, $i.e.$ $\Gamma_e=\Gamma_{\mu}$),
\beq \vec{\eta}=\eta(M_R, \Gamma_t, \Gamma_{\mu}, \Delta)~.
\label{veceta} \eeq Then chi-square estimator~\cite{minuit} can be
constructed as follows\footnote{Here for briefness, we only consider
the uncorrelated form of estimator. More complicated form with the
correlation between data taking into account could be referred to
Refs.~\cite{moxh03,moxh06,moxh03b}.} \beq
\chi^2=\sum\limits^{n_{pt}}_{i=1} \sum\limits^{h,e,\mu}_{f} \left(
\frac{n_f^i-N_f^i (\vec{\eta})}{\delta n_f^i}\right)^2~.
\label{chisqfm} \eeq Minimizing $\chi^2$ yields best values
(estimates) of the parameters wanted~\cite{minuit}. Usually Poisson
distribution is assumed for the data, then the relation $\delta
n_f^i=\sqrt{n_f^i}$ is always adopted for the statistical
uncertainty of the data.

As aforementioned, for physics analysis the luminosity is often
determined by some physics process with salient characteristic
topology and large cross section. As an example, for $\psp$ scan the
luminosity is calculated by $\EE$ event as follows \beq L^i =
\frac{n_e^C(W_i)}{A_{e}^{QED}(\vartheta) \sigma_{e}^C(W_i)}~.
\label{eqlum} \eeq However, as mentioned before,  among $\EE$ events
besides the contribution due to QED process there is also the
contribution from the resonance decay and interference.
To know $n_e^C(W_i)$,
resonance parameters must be determined first. To solve this
intertwist difficulty, an iteration method is adopted.

For the $j$-th iteration, the luminosity at energy $W_i$ is
calculated as follows:   \beq L^i_{(j)} =
\frac{n_e(W_i)}{\sigma_{e}^{j} (W_i)}~, \label{eqlum} \eeq where
$n_e(W_i)$ is the observed number of events of $\EE$ final state,
and $\sigma_{e}^{j} (W_i)$ is the observed cross section (with
superscript ``$obs$'' removed) calculated by Eq.~\eref{xctobsee} at
$j$-th iteration. As the first step in the fit, the $\sigma_{e}^{1}
(W_i)$ is obtained by a rough guess of the parameters, or the
previously measured ones, if they are available ($e.g.$ PDG values),
then calculated by Eq.~\eref{eqlum} is the $L^i_{(1)}$ which can be
used to work out the expected numbers of events for the processes
interested, such as $N^i_h$, $N^i_{\mu}$ (refer to
Eq.~\eref{expnmb}). Then utilize the estimator of Eq.~\eref{chisqfm}
to get the fit parameters
$$ \vec{\eta}_{(1)}=\eta(M_R^{(1)}, \Gamma_t^{(1)}, \Gamma_{\mu}^{(1)}, \Delta^{(1)})~. $$
At the next step of the fit,
%assuming $e$-$\mu$ universality, namely $B(\psp\to\EE)=B(\psp\to\MM)$,
the estimated resonance parameter values can be obtained directly
from the measured data, which in turn are used to acquire the
approximate total observed $\EE$ cross section $\sigma_{e}^{2}
(W_i)$  by Eq.~\eref{xctobsee}.

With the recalculated luminosity, the parameters
$$ \vec{\eta}_{(2)}=\eta(M_R^{(2)}, \Gamma_t^{(2)}, \Gamma_{\mu}^{(2)}, \Delta^{(2)})~. $$
are fitted again.
Such a recursive iteration is repeatly carried out until the
corrected values of $ L^i_{(j)} $ are converged in two successive
iterations, that is
$$ \left| \frac{L^i_{(j+1)}-L^i_{(j)}}{L^i_{(j+1)}}\right|< \alpha~.$$
Here $\alpha$ is the convergence precision which can be adjusted to
meet the need of scan fit.

\section{Summary}
For $\EE$ collision experiment, the approximate analytic formulas of
cross sections are presented for final states of inclusive hadron,
$e$-pair, and $\mu$-pair, where the experimental effects are taken
into account including initial radiative correction, vacuum
polarization, and energy spread. The experimentally observed cross
sections are also presented which take into account of the
acceptance of events and the contributions of the resonance, the
continuum, and the interference between them.

In the light of $e$-$\mu$ universality, the iteration technique is
adopted to figure out the cause-consequence interdependence between
the measurement of resonance parameter and determination of
luminosity. Such a kind of methods have been used successfully for
resonance parameters measurement of $\psp$ by BES
collaboration~\cite{bespspscan}.

\end{document}